\documentclass{PoS}

\usepackage{epsfig,multicol,multirow}
\usepackage{amsmath,subfigure,latexsym,amssymb}

\newcommand{\be}{\begin{equation}}
\newcommand{\ee}{\end{equation}}
\newcommand{\bea}{\begin{eqnarray}}
\newcommand{\eea}{\end{eqnarray}}
\newcommand{\med}{\frac{1}{2}}

\newcommand{\Yp}{\hat{Y}}
\newcommand{\calL}{{\mathcal{L}}}
\newcommand{\Tr}{{\rm Tr}}
\newcommand{\trm}{{ t}}

\newcommand{\real}{\mathbb{R}}
\newcommand{\field}{\bar{\Phi}}
\newcommand{\ccv}{\mathbf{c}}

\newcommand{\ijtp}{\emph{Int.\ J.\ Mod.\ Phys.}}
\newcommand{\cqg}{\emph{Class.\ and Quant.\ Grav.}}
\newcommand{\cmp}{\emph{Commun.\ Math.\ Phys.}} 
\newcommand{\ijmpa}{\emph{Int.\ J.\ Mod.\ Phys.} {\bf{A}}}

\newcommand{\jhep}{\emph{JHEP}}

\PoS{PoS(LAT2005)263}

\title{Field Theory Simulations on a Fuzzy Sphere ~ --- \\
an Alternative to the Lattice}

\ShortTitle{Field Theory Simulations on a Fuzzy Sphere - an Alternative to the
Lattice }

\author{\speaker{Julieta Medina}$^{~a,b}$ , 
Wolfgang Bietenholz$^{~c}$ , Frank Hofheinz$^{~c}$
and Denjoe O'Connor$^{~a,b} ~ $ 
\thanks{J.\ Medina would like to 
thank the {\it{Dublin Institute for Advanced Studies}} for hospitality 
and financial support. 
\hspace*{7mm} Preprint HU-EP-05/50} \\ 
  \ \\
 
  $^{a}$ Departamento de F{\rm\'\i}sica \\
  ~~ Centro de Investigaci\'on y de Estudio Avanzados del IPN \\
  ~~ Apdo. Postal 14-740 \\
  ~~ 07000 M\'exico D.F., M\'exico \\ 
  \ \\
  $^{b}$ {School of Theoretical Physics}  \\
  ~~ Dublin Institute for Advanced Studies \\
  ~~ 10 Burlington Road, Dublin 4, Ireland \\ 
  \ \\
  $^{c}$ Institut f\"{u}r Physik \\ 
  ~~ Humboldt Universit\"{a}t zu Berlin \\
  ~~ Newtonstr.\ 15, D-12489 Berlin, Germany \\
  \ \\
  E-mail: \email{julieta@synge.stp.dias.ie, bietenho@physik.hu-berlin.de, \\
\ \ \ \ \ \ hofheinz@physik.hu-berlin.de, denjoe@synge.stp.dias.ie}}

\abstract{We explore a new way to simulate quantum field theory, without
introducing a spatial lattice. As a pilot study
we apply this method to the 3d $\lambda \phi^4$ model. 
The regularisation consists of a fuzzy sphere with radius $R$ for the 
two spatial directions, plus a discrete Euclidean time. The 
fuzzy sphere approximates the algebra of functions of the sphere 
with a matrix algebra, and the 
scalar field is represented by a Hermitian $N \times N$ matrix at each 
time site. We evaluate the phase diagram, where we find a disordered phase 
and an ordered regime, which splits into phases of uniform and non-uniform 
order. We discuss the behaviour of the model in different limits of large 
$N$ and $R$, which lead to a commutative or to a non-commutative 
$\lambda \phi^4$ model in flat space.}

\FullConference{XXIIIrd International Symposium on Lattice Field Theory\\

                 25-30 July 2005\\

                 Trinity College, Dublin, Ireland}

\begin{document}

\section{The $\lambda \phi^4$ model on a fuzzy sphere}

In this work we deal with the {\em fuzzy sphere} formulation as a method
to discretise quantum field theory, and we apply it in numerical simulations
of the 3d $\lambda \phi^{4}$ model. In that scheme, the regularisation
of the spatial part of the action takes place in angular momentum space,
rather than coordinate space, hence no space lattice is involved.
Nevertheless we arrive at a finite set of degrees of freedom, which
allows us to study the model non-perturbatively.

In particular we are going to regularise the continuum action
\bea
S ( \phi ) & = & \int_{S^1} d t \ s ( \phi , t ) \ , \nonumber \\
  s ( \phi , t ) &=&  \int_{S^2}\left[ \med \phi(t,x)
  \left( \frac{\calL^2}{R^2}-\partial_t^2 \right)
  \phi(t,x) + \frac{m^2}{2}\phi^2(t,x)
  +\frac{\lambda}{4}\phi^4(t,x)         
  \right] R^2 d \Omega \ ,  \label{continous_action}
\eea
where $ d \Omega = \sin \theta d\theta d \phi$, and $S_1$ has
circumference $T$. The scalar field $\phi(t,x)$ depends on the Euclidean 
time $t$ and the space coordinates $x_i(\theta, \phi)$, with 
the constraint $ \sum_{i=1}^3 x_i^2=R^2 $.
$\calL_i$ are the angular momentum operators, and
$\calL^2= \sum_{i=1}^3 \calL_i^2$.

We first focus on $s ( \phi , t )$, 
the spatial part of this action.
As a discretisation we replace $S^{2}$ by a fuzzy sphere \cite{fuzzy}:
this means that the coordinates $x_i$ are replaced by
the coordinate operators $X_i = \frac{2R}{\sqrt{N^2-1}}
L_i$ \ ($i = 1,2,3$), where $L_{i}$ are the  $SU(2)$
generators in the $N$-dimensional irreducible representation.
These coordinate operators satisfy the constraint
$ \sum_{i=1}^3 X_i^2= R^2  \cdot 1 \!\!\! 1 \ $,
which corresponds to a matrix equation for a sphere.
The $X_i$ obey the commutation relation
\be
\left[X_i,X_j  \right]= i \epsilon_{ijk} 
\frac{2 R}{\sqrt{N^2-1}}X_k \ . \label{fuzzy_commutator}
\ee
At finite $N$ our coordinates describes a non-commutative geometry;
the sphere turns fuzzy.

As in the continuum, where $\phi$ can usually
be expressed as a polynomial in the coordinates $x_i$, its fuzzy 
counterpart $\Phi$ can be written as a polynomial in the coordinate operators.
This formulation is obtained by replacing $C^{\infty}(S^2)$, the
algebra of smooth functions on the sphere, by ${\tt Mat}_N$, a sequence of
matrix algebras of dimension $N$, where all positive integer
values of $N$ are permitted \cite{fuzzyphi4}.
Thus the scalar field is represented by a Hermitian matrix $\Phi$ 
of dimension $N$
(note that $\phi \in \real$ implies the Hermiticity of the matrix $\Phi$).

The differential operators ${\cal L}_i \, \cdot$ are
replaced by $\left[L_i, \cdot \, \right]$, and the integral over $S^2$ is
converted to the trace. The standard basis for functions on the sphere --- 
given by the spherical harmonics $\{Y_{lm} \}$ --- is replaced by the 
polarisation tensor basis $\{  \Yp_{lm} \}$, see for instance Ref.\ 
\cite{polten}. Let us summarise this set of substitutions:
\bea
x_i\in C^{\infty}(S^2)  \ \longrightarrow \ X_i \in {\tt Mat}_{N}  & , & \quad 
\phi(x)\in C^{\infty}(S^2) \ \longrightarrow \ \Phi \in {\tt Mat}_{N} \ , 
\label{replace_fuzzy_1} \\
{\cal L}_i \phi(x) \ \longrightarrow \ \left[ L_i, \Phi \right] & , & \quad
{\cal L}^2 \cdot \longrightarrow \hat {\cal L}^2 \cdot :=  \sum_{i=1}^3
\left[L_i, \left[L_i, \cdot  \right] \right] \ , \label{replace_fuzzy_5}
\label{replace_fuzzy_2} \\
\int_{S^2} \phi(x) d \Omega & \longrightarrow & \frac{4 \pi}{N} 
\Tr \left(\Phi \right) \ . \label{replace_fuzzy_3}
\eea

Rotations  on the fuzzy sphere are performed by using
an element $U$ of the $N$ dimensional unitary irreducible
representation of $SU(2)$.
A general element $U$ of this representation has
the form $U = \exp (i \omega_i L_i)$, with $\omega_i \in \real$.
The coordinate operators are then rotated as
$ U X_i U^{\dagger}= {\mathtt{R}}_{ij} X_j \, , \ \mathtt{R} \in SO(3) $,
and the field transforms as
$ \Phi \longrightarrow \Phi' = U \Phi U^{\dagger} $.

Implementing the substitutions 
(\ref{replace_fuzzy_1})-(\ref{replace_fuzzy_3}) in the action
(\ref{continous_action}), we obtain
\be
s ( \Phi (t)) = \frac{4 \pi R^2 }{N} 
\Tr \Big[ \frac{1}{2} \Phi(t) \left( \frac{\hat {\cal L}^2}{R^2}  
- \partial_{t}^2 \right) \Phi(t)  + \frac{m^2}{2} \Phi^2(t)
+\frac{\lambda}{4} \Phi^4(t) \Big] \ . \label{action_fuzzy_sphere}
\ee
This discretisation (at finite $N$) preserves the exact rotational symmetry 
of the continuum model, since any rotation on the sphere is allowed,
and action (\ref{action_fuzzy_sphere}) remains invariant.

To discretise the time direction we take a set of $N_{\trm}$
equidistant points, $T=N_{\trm} \Delta t $,
with periodic boundary conditions.
This yields the fully regularised action
\be
\hspace*{-2mm}
S \left[ \Phi\right] = \frac{4  \pi R^2 \Delta t}{N} \sum_{\trm=1}^{N_{\trm}} 
\Tr \Big[ \frac{1}{2R^2} \Phi\left(t\right)
\hat {\cal L}^2 \Phi\left(t\right) +  \med 
\left( \frac{\Phi(t+\Delta t) -\Phi(t)}{\Delta t}  \right)^2
+ \frac{m^2}{2} \Phi^2(t) +\frac{\lambda}{4} \Phi^4(t)
\Big] \ . \label{action_1}
\ee
We are interested in the limits $N\longrightarrow \infty$ and 
$N_{\trm}\longrightarrow \infty$, and for our simulations we fixed
$N=N_{\trm}$. 
One configuration $\Phi$ corresponds to a set of matrices
$\{ \Phi (t) \}$, for $\trm= \Delta t , \dots , N_{t} \Delta t $.

The formal expression for the Fourier decomposition of the field reads
\be
\Phi (t) = 
\sum_{l=0}^{\infty} \ \sum_{m = -l}^{l} \ \sum_{k \in Z \!\!\! Z} 
\, c_{lm}(k) \,
\exp \left( i \frac{2 \pi k t}{N_{\trm}} \right) \, \Yp_{lm} \ , \quad
c_{lm}(k) = \frac{4 \pi R^{2}}{N N_t} \sum_t \exp(-i \frac{2\pi kt}{N_t})
\Tr (\hat Y_{lm}^{\dag} \Phi(t))
\label{expansion_fourier}
\ee
In particular the temporal zero mode of $\Phi$ is given by
\be
\field = \frac{1}{N_{\trm}}\sum_{\trm=1}^{N_{\trm}}\Phi(t) = 
\ccv_{00} \Yp_{00}+\sum_{m=-1}^{1}\ccv_{1m}\Yp_{1m}+ \dots 
\label{averaged_field}
\ee
where
$ \ccv_{lm}:=\frac{1}{N_{\trm}}\sum_{\trm} c_{lm}(\trm).$
Our order parameters are based on the coefficients $\ccv_{lm}$,
\be
\varphi_0 := \vert \ccv_{00}  \vert \ , \qquad
\varphi_1 := \sqrt{ \sum_{m=-1}^{1} \vert c_{1,m} {\vert}^2 } \ ,
\label{phi_1}
\ee
and the corresponding susceptibilities.
We define $\varphi_{all}^2$ as the norm of the field $\field$,
\be
   \varphi_{all}^2:=\sum_{l,m} \vert \ccv_{lm} \vert^2=\frac{4 \pi}{N}
                \Tr\left( \field^2\right)= \varphi_0^2+ \varphi_1^2 +
\dots  \label{full_power_field}
\ee

\vspace*{-6mm}

\section{Numerical results}

\vspace*{-2mm}

A numerical study of the 2d version of this model was
presented in Ref.\ \cite{xmartin}. However, as an important 
qualitative difference, in that case the radius $R$ could be
absorbed in the couplings, whereas here it takes the r\^{o}le 
of an independent parameter.

We also recall that our formulation corresponds to a non-commutative space
on the regularised level. In analogy to previous studies of the
non-commutative $\lambda \phi^{4}$ model in flat spaces \cite{NCphi4}, and
to the 2d $\lambda \phi^{4}$ on a fuzzy sphere \cite{xmartin}, 
we observed three phases:
\begin{itemize}
 \item I : The disordered phase, characterised by
$\varphi_{all}^2\approx 0$, $\varphi_0 \approx0$, $\varphi_1 \approx0 , \ \dots$
 \item II : The uniform order, characterised by
$\varphi_{all}^2 \approx \varphi_0^2 > 0$, $\varphi_1\approx0$, $\dots$
 \item III : The non-uniform order, e.g.
 $\varphi_{all}^2 > 0$, 
$\varphi_0 \approx0$; $\varphi_1 > 0$, $\varphi_2 \approx0, \ \dots$
\end{itemize}

\begin{figure}[htbp]
\vspace*{-2mm}
\hspace*{-3mm}
\includegraphics[angle=270,width=0.5\textwidth]{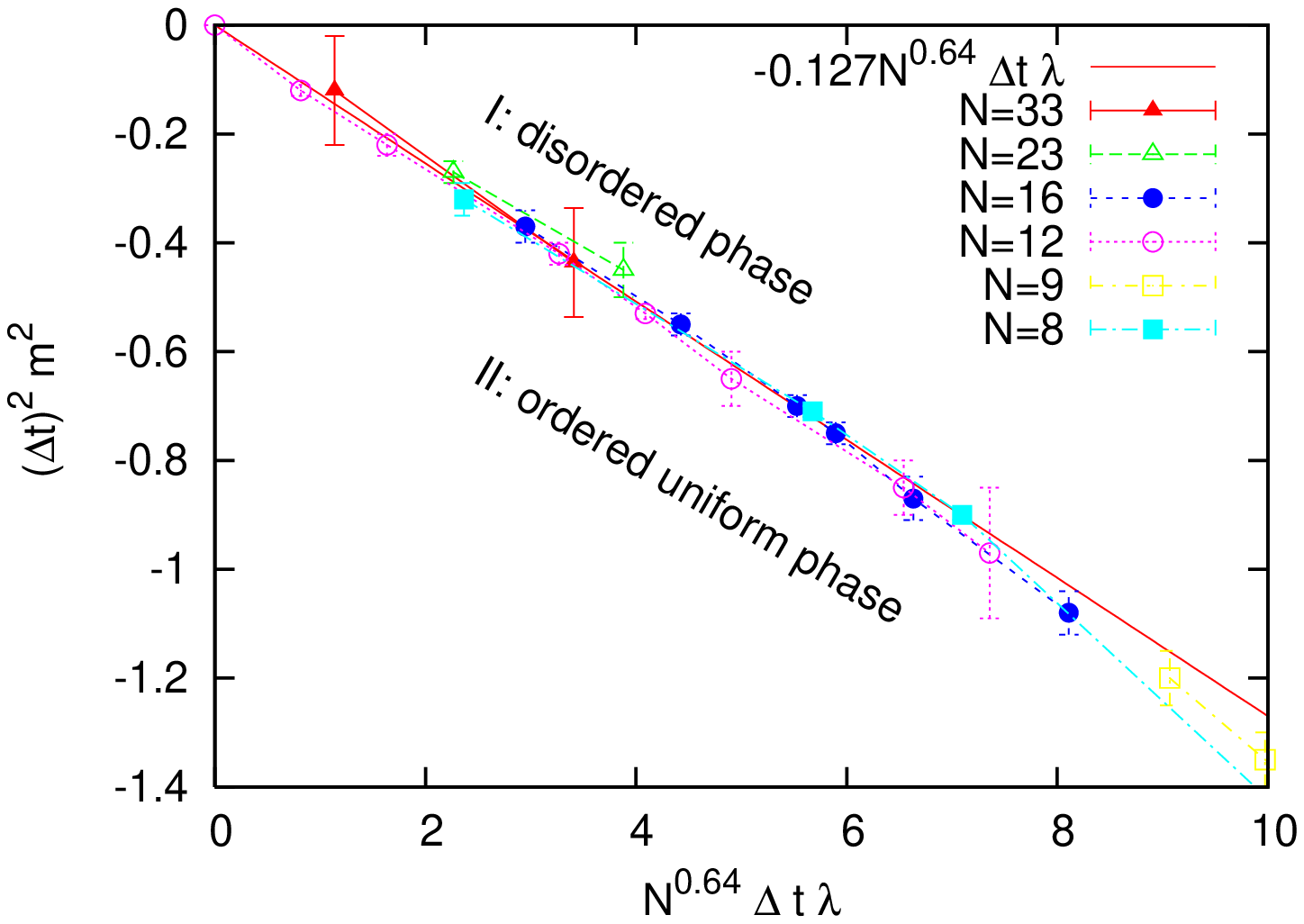}
\includegraphics[angle=270,width=0.5\textwidth]{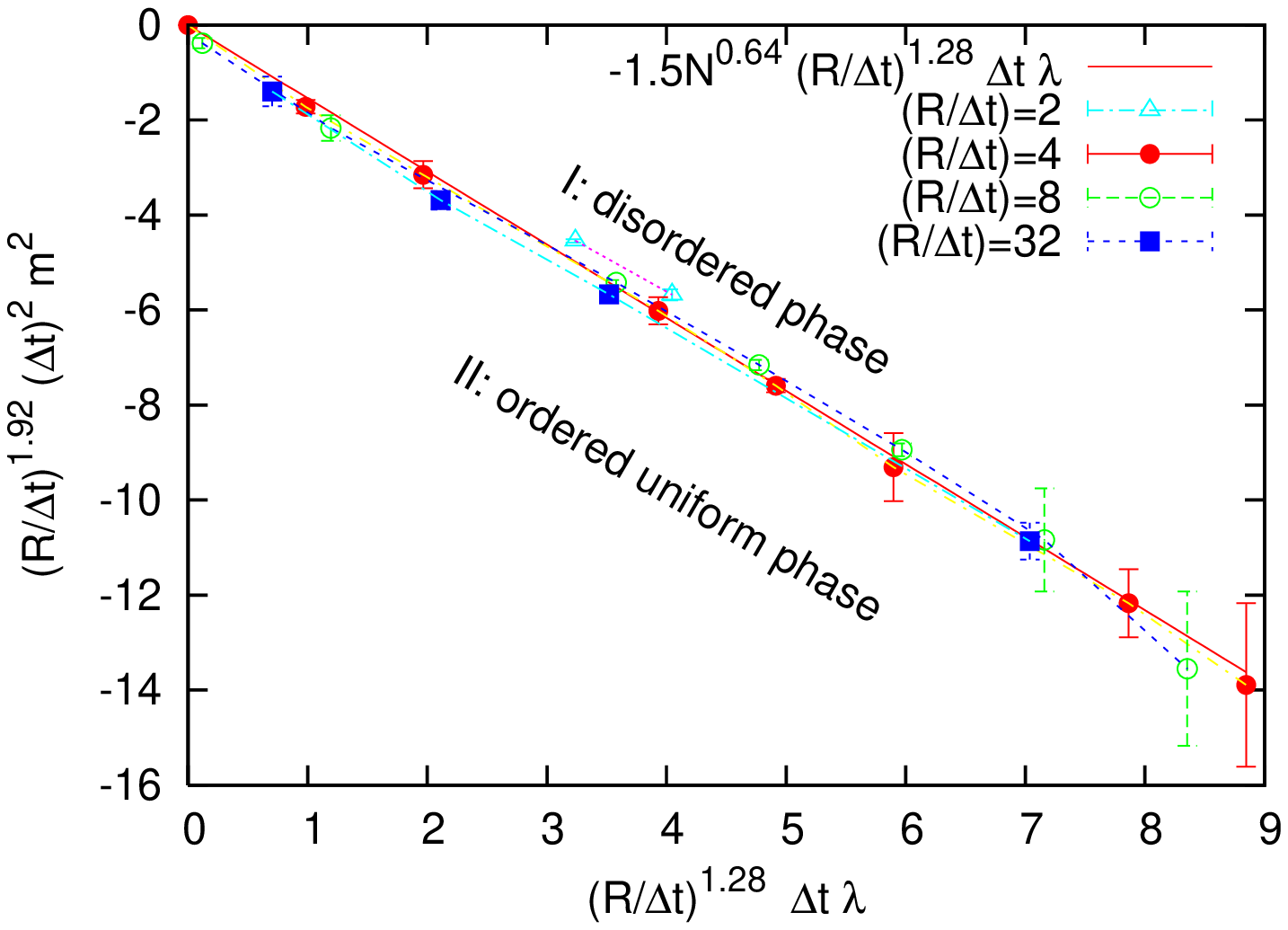}
\caption{Examples for the phase transition curves I-II: 
On the left we fix $\frac{R}{\Delta t}=4$ and vary $N$.
On the right we fix $N=12$ and and vary the ratio $\frac{R}{\Delta t}$. 
In both cases the transition lines stabilise for suitably scaled axes.}
\vspace*{-3mm}
\label{Fig1}
\end{figure}

\begin{figure}[htbp]
\vspace*{-4mm}
\hspace*{-3mm}
\includegraphics[angle=270,width=0.5\textwidth]{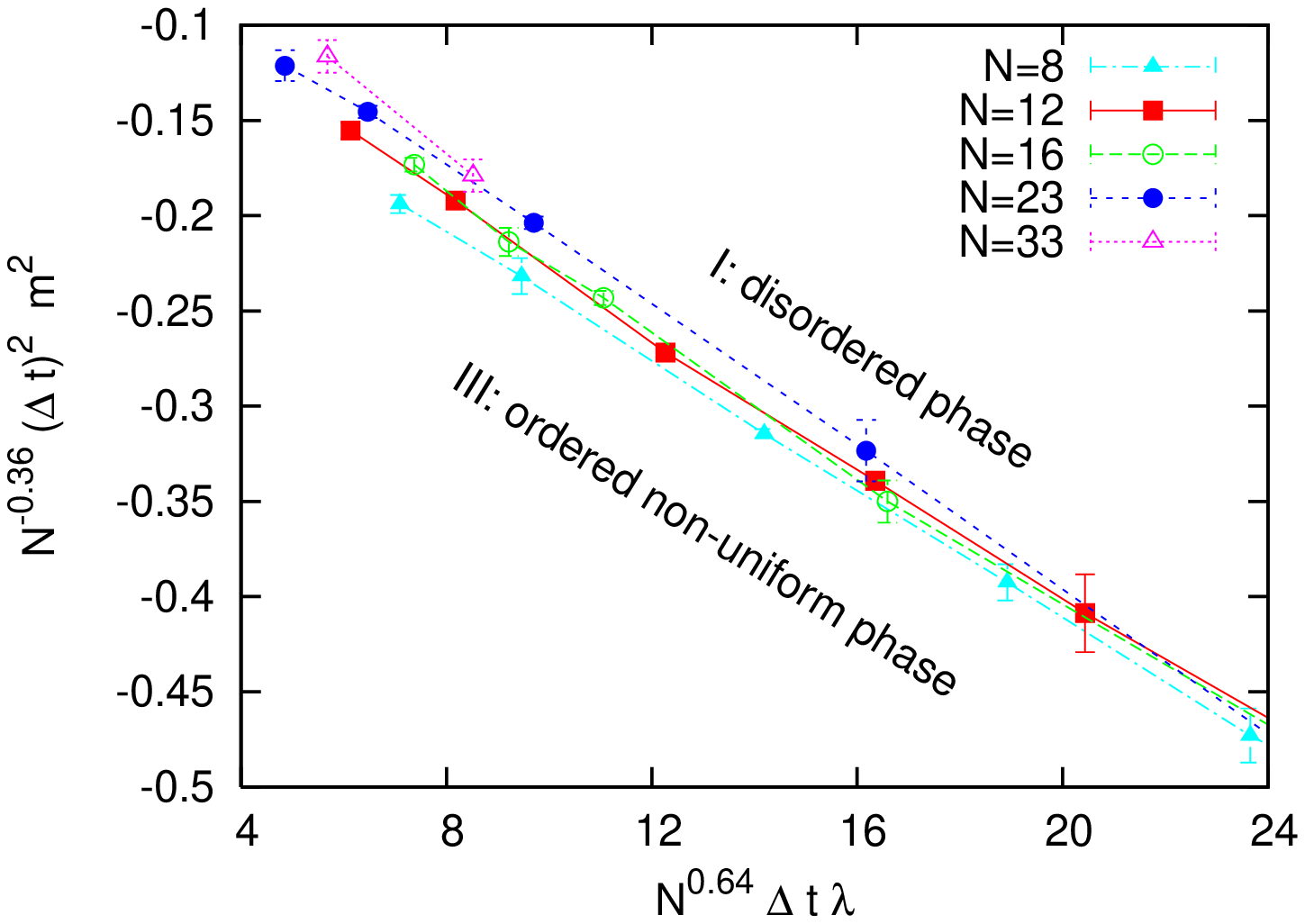}
\includegraphics[angle=270,width=0.5\textwidth]{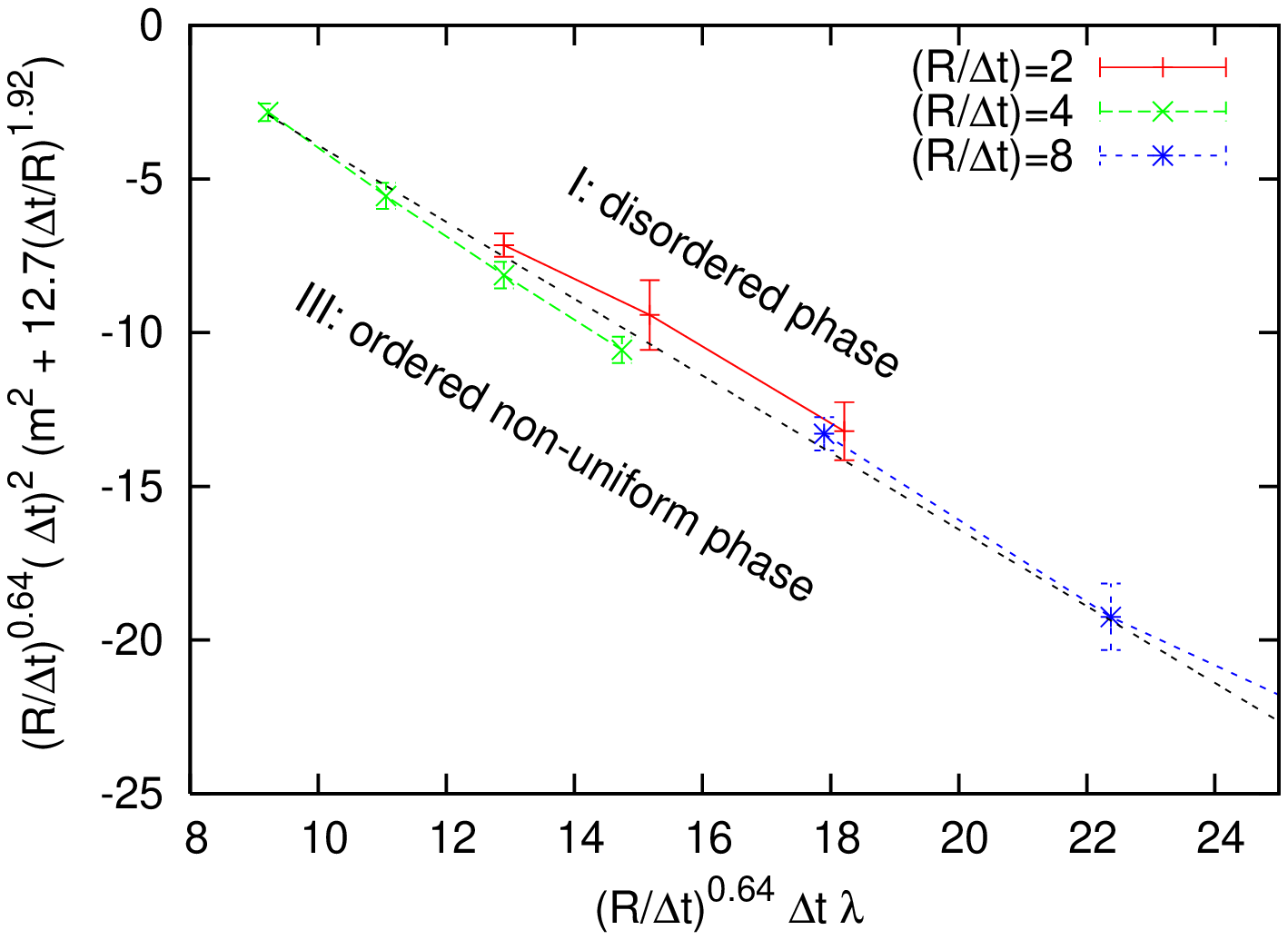}
\caption{Examples for the phase transition curves I-III: 
On the left we fix $\frac{R}{\Delta t}=8$ and vary $N$.
On the right we fix $N = 16$ and and vary the ratio $\frac{R}{\Delta t}$. 
Also here the transition lines stabilise for suitably scaled axes.}
\vspace*{-3mm}
\label{Fig2}
\end{figure}

\begin{figure}[htbp]
\vspace*{-4mm}
\hspace*{27mm}
\includegraphics[angle=270,width=0.53\textwidth]{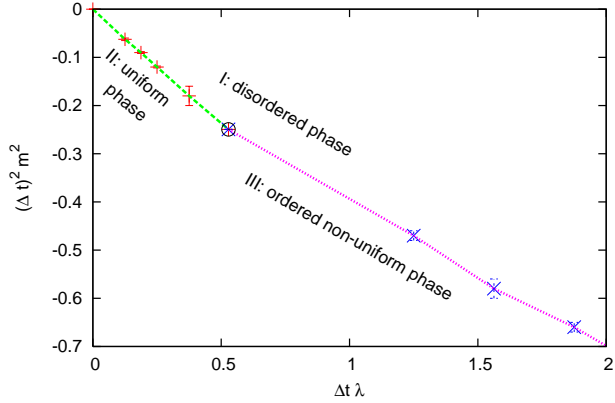}
\caption{The phase diagram for 
$N=16$, $\frac{R}{\Delta t}=4$. We show linear fits for both transition 
curves.  The extrapolation of those lines allow us to identify the triple 
point.} 
\vspace*{-3mm}
\label{Fig3}
\end{figure}

\noindent
Figures \ref{Fig1} to \ref{Fig3}
give an overview of our numerical results.

The triple point $ \left( \lambda_T,  m^2_T \right)$ is fixed by the 
intersection of the transition curves $I-II$ and $I-III$.
We focus on its behaviour since it determines 
which phases survive under different limits. 
The emerging triple point expression reads
\be 
  \left( \Delta t \lambda_T, \Delta t^2 m^2_T \right) = \left(
(41.91\pm30)\left(\frac{\Delta t^2}{ N R^2}\right)^{\gamma}, 
- (12.7\pm1) \left( \frac{\Delta t}{R}\right)^{3 \gamma '}  \right) \ ,
\label{f-point-R-function-N}
\ee
where our numerical results are consistent with
$\gamma = \gamma' =0.64 \pm 0.2$.

\vspace*{-2mm}

\section{Behaviour of the model under different limits}

\vspace*{-1mm}

For the temporal part of the model we set
$ \Delta t = \frac{1}{N^{\kappa}}$, $\kappa \in (0,1) $, 
hence the time extent $T$ amounts to $N^{1-\kappa}\longrightarrow \infty$.

\vspace{-3mm}

\begin{itemize}
\item
Regarding the spatial part of the model, the non-commutative 
{\em Moyal plane limit} $\real_{\Theta}^2$ can be accessed if
$ R^2=\frac{N \Theta}{2}$, for $\Theta$ fixed and 
$ R, \, N \longrightarrow \infty$ \ \cite{steinacker}. 

\vspace{-2mm}

\item
For the {\em commutative flat limit}, $\real^2$, we take 
$R \propto N^{\med(1-\epsilon )}$ with  $\epsilon \in (0,1)$, and 
$ N \longrightarrow \infty$.

\vspace{-2mm}

\item
The {\em commutative sphere limit} arises when $R$ is fixed 
and $N \longrightarrow \infty$, since  $ C^{\infty}(S^2)$ is recovered.

\end{itemize}

\vspace{-3mm}

\noindent
We summarise all cases by setting $R=N^{\beta}$ with
$\beta= \med (1-\epsilon )$, $\epsilon \in [0,1]$.
Then eq.\ (\ref{f-point-R-function-N}) takes the form
\be 
\left(  \lambda_T,  m^2_T \right) = \left(
41.91 N^{\kappa (1-2 \gamma)-\gamma(1+2 \beta )}, 
-12.7 N^{2\kappa -3 \gamma ' ( \beta + \kappa)}  \right) \ .
\label{f3-point-R-function-N}
\ee
Although the error on $\gamma$ resp.\ $\gamma'$ is sizable,
the exponent in $\lambda_T$ in eq.\ (\ref{f3-point-R-function-N}) 
seems to be clearly negative. 
Therefore $\lambda_T \longrightarrow 0$ as 
$N\longrightarrow \infty$, which indicates the disappearance of
the uniform order phase.

As a particular case we consider $\kappa=\med$,
with the tri-critical action
\bea
S_T ( \Phi ) & \approx & \sum_{\trm=1}^{N} 
\Tr \Big[ \frac{2 \pi}{N} \Phi\left(t\right)
\hat {\cal L}^2 \Phi\left(t\right) +  2 \pi  N^{2\beta}
\left[ \Phi(t+\Delta t) -\Phi(t)  \right] \nonumber \\ & &
\qquad -\frac{25.4 \pi}{N} \Phi^2(t)
+ \frac{41.9\pi }{N} N^{\gamma(\beta-\med) } \Phi^4(t) \Big] \ .
\label{action_triple-v2}
\eea
For large $N$ the leading contribution to eq.\
(\ref{action_triple-v2}) is the temporal kinetic term, while 
the contribution from the fuzzy kinetic term is negligible. Hence the
uniform order phase disappears in this limit, which leads to
a simplified model that was analysed in Refs.\ \cite{BEynard}.

Finally we compare our results to those obtained in the last work
quoted in Ref.\ \cite{NCphi4}. 
If we set $\Delta t=1$ and $R=N$, we arrive at a similar behaviour of the triple 
point, namely $\left(N^2 \lambda_T,N^2 m^2_T \right) \approx const$. 
The suitable action is given by eq.\ (\ref{action_triple-v2}) at
$\beta = 1/2$.
Again the leading contribution is due to the temporal kinetic term.

\vspace*{-2mm}

\section{Conclusions}

\vspace*{-1mm}

We presented a numerical study of the $\lambda \Phi^4$ model  on the 
3 dimensional Euclidean space where we combined two schemes of discretisation.
As in related models studied previously \cite{xmartin,NCphi4}, 
we identified the existence of three phases, one of which is unknown
in the $\lambda \phi^{4}$ model in the continuous commutative space.
The fate of these phases under various limits will be discussed in more
detail in Ref.\ \cite{prep}.

At this point, we just repeat that the triple point scales to zero
in the limit $N\rightarrow\infty$.
Hence this simple model cannot capture the Ising universality class in
this limit. This is not surprising, given the perturbative results of
Refs.\ \cite{fuzzyphi4} and \cite{steinacker}, 
where it was observed (in the commutative limit)
that though the non-planar diagrams have the same divergence 
at $N \rightarrow \infty$
as the planar diagrams, the difference of the two diagrams is finite and
non-local. 

In any case, to maintain the uniform phase, it is necessary to reinforce
the fuzzy kinetic term. This is achieved most simply by adding a higher
derivative contribution which will guarantee that all diagrams are
convergent in the large $N$ limit. For the current model this would
correspond to adding a term 
$\Phi (\hat {\cal L}^2)^2 \Phi / (\Lambda^2 R^4)$ inside the
trace in eq.\ (\ref{action_1}) (where $\Lambda$ is a momentum cutoff). 
By an appropriate scaling of $\Lambda$ it should 
be possible to send the triple point to infinity as 
$N \rightarrow \infty$.\\


This pilot study reveals that the fuzzy sphere
formulation does indeed enable numerical simulations without
the requirement of a spatial lattice. Its virtue as a discretisation
scheme is that it preserves certain symmetries exactly,
which are explicitly broken on the lattice.
We hope for that virtue to become powerful in particular
in supersymmetric models.\\

%


\vspace*{-2mm}

\noindent
{\small {\bf Acknowledgements } \ \  
We are indebted to A.\ Balachandran, B.\ Dolan, X.\ Martin, J.\ Nishimura,
M. Panero, P.\ Pre\v{s}najder, H.\ Steinacker, J.\ Volkholz and B.\ Ydri
for inspiring discussions. 
This work was supported in part by the ``Deutsche Forschungsgemeinschaft'' 
(DFG).}

\end{document}